%% file: 00_main.tex
\RequirePackage{fix-cm}
\documentclass[twocolumn]{svjour3}          
\smartqed  

\usepackage{microtype}
\usepackage{mathptmx}      
\usepackage{amsmath, amssymb}
\usepackage{graphicx}
\usepackage{caption, subcaption}
\usepackage{threeparttable, booktabs}
\usepackage{cite}
\usepackage[misc]{ifsym}

\usepackage{xurl}
\usepackage{hyperref}

\usepackage{booktabs}
\usepackage{multirow}

\usepackage{xcolor}
\newcommand{\pos}{\textcolor{green!70!black}{Positive}}
\newcommand{\nega}{\textcolor{red}{Negative}}

\usepackage[linesnumbered,ruled,vlined]{algorithm2e}
\usepackage{algpseudocode}
\newtheorem{constraint}{Constraint}

\journalname{Preprint}
\makeatletter
\def\makeheadbox{\relax}
\makeatother
\begin{document}

\title{LMG Index: A Robust and Efficient Learned Index Framework for Multi-Dimensional Performance Balance}


\author{Yuzhen Chen \textsuperscript{1}\and
        Bin Yao \textsuperscript{1}
}


\institute{
        \Letter Bin Yao \at
        \email{yaobin@cs.sjtu.edu.cn}   
        \and
        Yuzhen Chen \at
        \email{chenyz-sjtu@sjtu.edu.cn}
        \at
        {1} Shanghai Jiao Tong University, Shanghai, China 
}

\date{Received: date / Accepted: date}

\maketitle

\begin{abstract}
Index structures are fundamental for efficient query processing on large-scale datasets. Learned indexes model the indexing process as a prediction problem to overcome the inherent trade-offs of traditional indexes. However, most existing learned indexes optimize only for limited objectives like query latency or space usage, neglecting other practical evaluation dimensions such as update efficiency and stability. Moreover, many learned indexes rely on assumptions about data distributions or workloads, lacking theoretical guarantees when facing unknown or evolving scenarios, which limits their generality in real-world systems.

In this paper, we propose LMG, a robust and efficient learned index framework designed for multi-dimensional performance balance. LMG integrates a decoupled routing structure with theoretical $O(1)$ complexity for fixed key types and an optimal error threshold training algorithm that approaches $O(1)$ overhead in practice. Furthermore, the framework enhances query performance by optimizing gap allocation. Extensive evaluations show that our framework achieves competitive or leading performance across all key evaluation dimensions, including bulk loading (up to 7.55$\times$ faster), point queries (up to 1.68$\times$ faster), range queries (up to 11.41$\times$ faster), and mixed read-write throughput (up to 3.50$\times$ faster). Furthermore, LMG ensures robust long-term stability and high space efficiency (up to 6.26$\times$ smaller footprint). These results demonstrate that LMG significantly mitigates the multi-dimensional performance trade-offs often observed in state-of-the-art approaches, offering a balanced and efficient framework.

\keywords{Learned index \and Robust framework \and Data access}
\end{abstract}

\input{01_intro}

\input{02_background_motivation}

\section{LMIndex Overview}
\label{sec:overview}
\begin{figure}[h]
    \centering
    \includegraphics[width=1\linewidth]{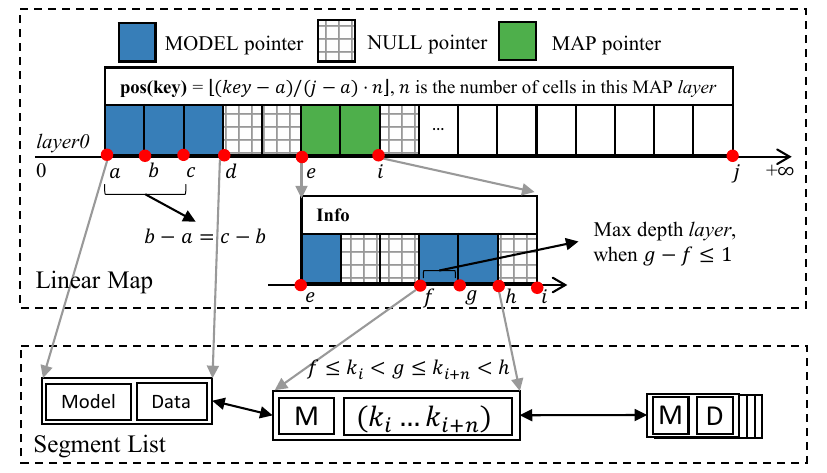}
    \caption{Main structure of LMIndex. Linear Map is its segment indexing structure. Each cell within a layer represents a fixed key interval, so the depth of the layer is maximized when the interval is $\le 1$. Precise routing to the covering cell is achieved through layer parameters, such as key range and cell count.}
    \label{fig:main-structure}
\end{figure}


LMIndex is an in-memory, space- and time-efficient versatile learned index paradigm designed to adapt to unknown or dynamic workloads and data
distributions. Its key idea is straightforward: use a lightweight yet effective structure that preserves high performance across diverse workloads without complex online tuning. To achieve this, LMIndex adopts a simple two layer architecture (Figure \ref{fig:main-structure}).

At the top layer, Linear Map (LM) efficiently identifies the target segment. As an auxiliary structure, LM neither requires online parameter tuning to maintain stability nor relies on costly model fitting. Instead, it structurally captures the key range distribution of the segments (i.e. $[key_{min}, key_{max}]$ for each segment). And it can locate the bottom segment in $O(1)$ time complexity when the key type is determined.

The bottom layer consists of an ordered sequence of segments, each primarily containing a linear model and the data layout. The linear model includes a range (max/min key), parameters, and an error threshold. To handle dynamic workloads, the segment implements a flexible update absorbing function, where incoming updates are first buffered locally by the segment. This buffering mechanism offers multiple architectural choices: it can employ an out-of-place structure, utilizing an additional secondary index like a red-black tree or a linked list, or an in-place structure, such as building child nodes similar to \cite{lipp21}. Within our framework, LMIndex employs a dense array coupled with an out-of-place buffer, whereas its variant, LMG, utilizes a gapped array as an in-place buffer. For LMG, we specifically employ an optimized gap allocation strategy. By distributing the reserved gaps more evenly across the array, this strategy reduces the average error search length and improves operational efficiency.



This architecture is motivated by the goal of fully leveraging the inherent structural quality of the LM. Consequently, modest augmentations yield meaningful improvements, enabling the framework to strike a robust balance and secure advantages across multiple performance dimensions:

\textbf{Bulk load.} The bulk load consists of the construction of bottom layer and top layer. The bottom layer is constructed in $O(n)$ time by integrating a linear segmentation algorithm with the efficient model training algorithm, OETA. In the top layer, LM avoids complex approximation fitting, thus ensuring that the construction time overhead is $O(m)$ ($m$ is the number of segments, less than the data size). So the total complexity of bulk load is $O(n)$.

\textbf{Updates.} To handle dynamic workloads, the framework temporarily buffers incoming modifications locally within the segment before triggering a Structural Modification Operation (SMO) upon reaching a predefined capacity. At the bottom layer, this SMO functions similarly to localized bulk loading, retraining exclusively on the affected segment data to preserve error bounds. Meanwhile, at the top layer LM, the SMO incurs a theoretical $O(1)$ cost, adding only a marginal overhead to the overall update time.

\textbf{Stability.} Many learned indexes degrade after updates because structural changes disrupt query performance. LMIndex maintains stability by ensuring that the post-update structure resembles that of a freshly bulk-loaded index on the current data.

\textbf{Point and range queries.} Queries are decomposed into a top layer LM lookup and linear model prediction with error correction at the bottom layer. LM and OETA together ensure consistently low latency for both steps.

\textbf{Space usage.} Due to the optimal segmentation and compact structure, reducing space usage at the bottom layer is less effective. Thus, optimization focuses on the top layer LM, whose storage depends on the key range distribution of the segments. On real-world datasets, LM accounts for only about 4\% of the total index size in average. (see Section \ref{sec:lm_eval})


\subsection{Basic Operations}
\subsubsection{Point Queries and Range Queries}
For a point query with a search key $q$, the lookup process first traverses the top layer structure LM to locate the target segment, followed by a prediction and correction search within that segment. The LM is composed of flat pointer arrays (referred to as map layers), where each layer is composed of several cells. As illustrated in Figure \ref{fig:main-structure}, the key range of \textit{layer0} is $[a, j]$ and it contains $n$ cells. By simply computing $pos(key) = \lfloor (key - a)(j - a) \cdot n \rfloor$, one can precisely locate the corresponding cell for a given key. Each cell contains one of different types of Pointers: a MODEL pointer directs the query to the target segment; a NULL pointer terminates the search indicating the key is absent; and a MAP pointer advances the traversal to the next map layer. For a fixed key type, the routing complexity of LM is highly efficient and bounded by $O(1)$ since the maximum layer depth is finite (details in Section \ref{sec:lm}).

Within the targeted segment, LMIndex utilizes the linear model to predict the physical position of the query key $q$ within the data layout. If the initial prediction deviates from the exact location, a bounded binary search is executed within the segment's predefined error threshold. Upon locating a matching, non-deleted key, the corresponding record is returned. Otherwise, the query execution path falls back to probe the local buffering mechanism to resolve any recently inserted or modified records.

For range queries, they are initiated by locating the first key greater than or equal to the predefined lower bound, followed by a sequential scan that terminates upon exceeding the upper bound. As illustrated in Figure \ref{fig:main-structure}, contiguous segments are sequentially linked via bidirectional pointers, facilitating efficient cross-segment traversals while naturally bypassing invalidated or deleted keys. Notably, when a segment employs an out-of-place buffering mechanism, the framework dynamically merges keys residing in the primary data array with the ordered elements held in the secondary local buffer during this traversal. While this on-the-fly merging inevitably introduces a slight latency overhead, its theoretical time complexity remains $O(K)$ (where $K$ is the number of retrieved keys). Furthermore, to optimize execution during extensive range queries, the framework proactively probes the required capacity and pre-allocates contiguous memory prior to the scan, effectively eliminating the computational overhead of repeated dynamic resizing.

In detail, we explore the query mechanics of LM in Section \ref{sec:lm}. For the ``last-mile'' search within the segment, we demonstrate in Section \ref{sec:single-model} how binary search outperforms exponential search in complex distributions dataset, while OETA narrows the segment's error threshold so that binary search performs on par with exponential search under near-linear distributions.


\subsubsection{Update}
\label{sec:update}
The insertion process initiates by routing the new key to its target segment. If the corresponding physical slot is occupied by a logically deleted record, the new key is directly inserted in-place. Conversely, the framework delegates the insertion to the segment's local buffering mechanism. To preserve the efficiency of the sequential scans, this local buffer maintains all newly absorbed elements in an ordered state. The maximum capacity of this buffer serves as a configurable parameter. Once the predefined capacity is saturated, the segment merges the buffered records and triggers a local Structure Modification Operation (SMO) to retrain the model.


For retraining, LMIndex employs OETA to train segment parameters. While it shares the same Big-O complexity as Least Squares (LS), its computational overhead is typically significantly lower in practice (see Section \ref{sec:oeta}). Prior to training, we apply the segmentation algorithm from \cite{pgm20} to partition the data. This process may yield a residual segment at the tail containing fewer than $2\varepsilon$ ($\varepsilon$ is the global error threshold) keys. To avoid its negative impact on lookup efficiency, we first attempt to merge the residual keys into the next segment, provided the resulting error remains within the $\varepsilon$-constraint. If the merge fails, they are jointly retrained. This residual-handling strategy may trigger cascading retraining, but this guarantees optimal segmentation, ensuring stability even in heavy write workloads. Further optimizations for this issue are presented in LMG (Section \ref{sec:lmg}).


For deletions, if a matching key is found within a segment, LMIndex marks the key with a tombstone. In-place deletion ensures that deletion efficiency is generally similar to querying. To maintain stability, LMIndex reallocates data after deletion: when the number of valid keys within a segment falls below global $\varepsilon$, the largest key from the preceding segment is moved and placed in front of the segment. If this insertion does not satisfy the $\varepsilon$-constraint, or if the number of valid keys in the previous segment is less than $\varepsilon$ + 1, the current and the previous segment are merged and retrained.


When insertions or deletions shift a segment’s key range, LM is updated to maintain segment routing. Because these changes are confined to affected segment boundaries, the update cost is negligible compared to retraining. Section~\ref{sec:lm} elaborates on LM’s update handling in detail.

\subsubsection{Bulk Load}

Bulk load consists of splitting the data, training the model, and building the segment index. First, LMIndex splits data using the algorithm in \cite{pgm20}. Given a global error threshold $\varepsilon$, it ensures each segment admits at least one linear function fitting all points within error bound $\varepsilon$.


Secondly, to further improve local accuracy, we introduce OETA, which efficiently computes the minimum error threshold and corresponding model parameters for each segment. OETA enhances local adaptivity of models, enabling smaller correction ranges. For instance, if global $\varepsilon = 64$ but the optimal threshold for a segment is $2$, correcting within that segment requires only $\log_2(2) = 1$ steps, i.e., a $\log_2(64)/\log_2(2) = 6\times$ reduction in correction steps when using binary search.



Finally, segments are indexed by the LM structure. LM is a shallow, hierarchical structure with theoretical construction cost $O(m)$, where $m$ is the number of segments. Larger range segments are placed in higher layers with coarser granularity, while smaller ranges descend into finer levels, allowing LM to adapt to segment range distribution. Detailed LM construction procedures are discussed in Section~\ref{sec:lm}. Total complexity of bulk loading is $O(n)$.

\subsection{LMIndex with Gaps (LMG)}
\label{sec:lmg}
LMG is a variant of LMIndex tailored to raise update throughput while preserving query efficiency. The primary structural modification in LMG is it replaces compact array with gapped array. And we introduce an expansion rate $\tau$ to control the additional space allocated for gaps.

Allocating gaps in a gapped array is usually implemented by inserting keys. Standard model-based insertion \cite{alex20} places a new key at its predicted position $f(k)$ or the nearest available empty slot. However, this approach accumulates localized errors under data skew. Formally, we define a contiguous sequence of inserted keys $K = \langle k_1, \dots, k_s \rangle$ as a \textit{many-to-one} subset if the prediction model maps them to the exact same position, i.e., $\forall k_j \in K, f(k_j) = C$, where $s = |K|$. Sequential insertion of $K$ forces the final key $k_s$ to position $C + s - 1$, yielding a lower-bound maximum error of $s-1$.

To mitigate this effect, LMG introduces a reserved-space insertion strategy that calculates an optimized starting physical index $P_{start}$ for the first key $k_1$. The allocation is strictly constrained by two spatial boundaries. Let $pos(x)$ denote the allocated physical index of key $x$. The lower bound $C_f$ is defined as $C_f = \max_{k'<k_1} (pos(k')) + 1$. Symmetrically, the upper bound $C_b$ is defined as $C_b = \min(f(k_{s+1}), N)$, where $N$ is the array capacity.

Given the predicted position $C$, subset size $s$, segment error threshold $\varepsilon_i$, and the spatial bounds $[C_f, C_b]$, the strategy dynamically determines $P_{start}$ through the following piecewise function:
$$ P_{start} = \begin{cases}
C_f & \text{if } C - C_f \le s/2 \\
\max(C_f, C - \varepsilon_i, C_b - s) & \text{if } C_b - C \le s/2 \\
C - s/2 & \text{otherwise}
\end{cases} $$
Following $k_1$, the remaining keys $k_j \in K$ ($2 \le j \le s$) are inserted contiguously at physical indices $pos(k_j) = P_{start} + j - 1$.

Functionally, this formal allocation aligns dense many-to-one subsets symmetrically to minimize localized maximum errors, while maintaining the same behavior as model-based insertion for other keys.




For updates, LMG employs a gapped array as its in-place buffering mechanism. While this in-place design introduces additional space overhead compared to out-of-place alternatives, it removes the overhead associated with cross-structure data access and on-the-fly merging during queries. During an insertion, the framework identifies the target slot and shifts the nearest available gap to accommodate the new key. To bound insertion latency, gap movement is restricted to a constant multiple of $\varepsilon$ (e.g., $8\varepsilon$). Furthermore, to prevent the cascading retraining issue discussed in Section \ref{sec:update}, LMG merges any residual tail segment shorter than $2\varepsilon$ into the preceding segment. Although this method slightly relaxes the $\varepsilon$-constraint, it effectively avoids cascading retraining overhead and segment fragmentation.

To guarantee semantic correctness of search despite potential update, LMG adopts a binary–then–exponential hybrid search strategy. It first performs a binary search bounded by the segment's error threshold, and falling back to exponential search if the target position is not located. As show in Section \ref{sec:choose-correction}, this design prioritizes binary search because it's more effective on complex data distributions.

\section{Linear Map}
\label{sec:lm}
In learned indexes based on piecewise linear models, the major overhead of queries typically stems from locating the correct segment and performing last-mile correction within it. Linear Map (LM) focuses on segment localization, achieving theoretical $O(1)$ query and update complexity. Its space complexity is $O(m)$, where $m$ is the number of segments, and the actual memory footprint adapts to the distribution of segment ranges.
\begin{figure}[h]
    \centering
    \includegraphics[width=\linewidth]{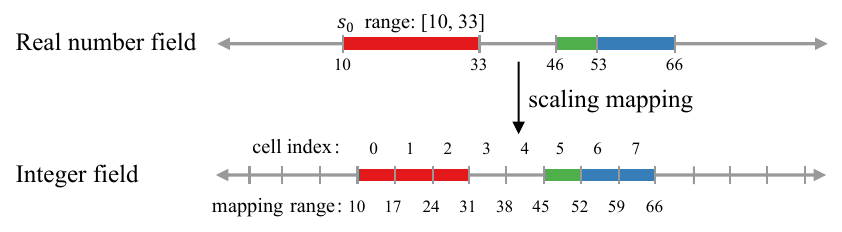}
    \caption{Key idea of LM: Mapping segment range from the real number field to the integer field, ensuring each segment corresponds to at least one integer cell.}
    \label{fig:lm-mapping}
\end{figure}

\subsection{Single-layer LM}
\label{sec:sl-lm}
Let $S = \{s_1, s_2, \dots, s_m\}$ denote an ordered set of non-overlapping segments, where each segment $s_i$ corresponds to the range of keys $[min_i,max_i]$ with width $R(s_i) = max_i - min_i$. These segments can be naturally viewed as intervals over the real number field, as illustrated in Figure \ref{fig:lm-mapping}.

To enable direct lookup without probing, Linear Map projects the continuous key range onto a flat indexable array, referred to as the \textit{layer}. Each cell in the \textit{layer} represents a fixed-length subrange of the key domain and stores either a pointer MODEL to the segment that covers that subrange or NULL if no segment overlaps it.

To ensure that every segment occupies at least one cell, LM guarantees minimum resolution over the key range $[min_1, max_m]$. Let $R(layer) = max_m - min_1$ and $|layer|$ be the number of cells in layer, the minimum resolution is:
$$
\min |layer| = \left\lceil R(layer) / \min R(s_i) \right\rceil
$$

LM adopts this minimal resolution to optimize space usage. Under this resolution, LM applies a scaling mapping formula (SMF) to directly compute the cell index for a key:
$$
\text{SMF}(key) = \left\lfloor (key - min_1) / R(layer) \cdot |layer| \right\rfloor
$$

We next show that SMF guarantees that the endpoints of every segment map to different cells when $|layer|$ equals the minimum required resolution. Let $d = \min R(s_i)$, and consider a segment range $[k_a, k_a + d]$. Then:
$$
\begin{aligned}
&\text{SMF}(k_a + d) \geq \left\lfloor (k_a - min_1) / R(layer) \cdot \left\lceil R(layer) / d \right\rceil\right\rfloor + 1 \\
&\text{SMF}(k_a + d) \le \left\lfloor (k_a - min_1) / R(layer) \cdot \left\lceil R(layer) / d \right\rceil \right\rfloor + 2
\end{aligned}
$$

Since $\text{SMF}(k_a) = \left\lfloor (k_a - min_1) / R(layer) \cdot \left\lceil R(layer) / d \right\rceil\right\rfloor$, $\text{SMF}$ $(k_a + d)$ is bounded by $[\text{SMF}(k_a) + 1, \text{SMF}(k_a) + 2]$, ensuring that each segment occupies at least one distinct layer cell and preventing boundary collapse.

Importantly, SMF is used for both queries and updates. Although it is a linear function, it differs from linear predictive models in that it requires no training. All parameters are derived directly from the known properties of the segment set $S$, enabling it to dynamically adapt to changes in the segment layout.

To construct a single-layer LM, the \textit{layer} is first initialized with NULL pointers. For each segment $s_i$, its range $[min_i,max_i]$ is projected to the cell range $[\text{SMF}(min_i), \text{SMF}(max_i)]$, and each cell in this range is filled with a MODEL pointer link to $s_i$. When adjacent segments $s_j$ and $s_{j+1}$ share a boundary cell, i.e., $\text{SMF}(max_j) = \text{SMF}(min_{j+1})$, LM assigns this cell to $s_{j+1}$. This minimum-first boundary rule consistently resolves ties without ambiguity and preserves correctness.

For point queries, SMF computes the target cell in $O(1)$ time. Every key is deterministically mapped to its corresponding cell with perfect accuracy. In most cases, the cell contains the correct segment pointer. In boundary cases where a key near the tail of segment $s_j$ falls into the first cell of $s_{j+1}$, LM performs a single fallback step to $s_j$ via the segment list, ensuring correctness with constant overhead.

For updates, LM locates the affected cells via SMF and replaces them with NULL or new MODEL pointers. The update cost is proportional to the number of cells traversed.

\subsection{Multi-layer LM}
\label{sec:ml-lm}
The single-layer LM represents an conceptually unbounded flat array, with space complexity $O(\min |layer|)$. Since the theoretical lower bound for indexing $m$ segments is $O(m)$, the single-layer LM amplifies:
$$
O(\min |layer|/m) = O\left( (R(layer)/m) / \min R(s_i) \right)
$$

When the distributions is skewed or non-linear, $\min R(s_i)$ may be orders of magnitude smaller than the average segment width $R(layer)/m$, resulting in excessive space usage, poorer cache locality, and higher average update cost. Moreover, future insertions that further reduce $\min R(s_i)$ force a full rebuild of the single-layer LM, incurring considerable overhead.

To avoid these limitations, we introduce the multi-layer LM, which bounds each \textit{layer} size by a constant $\mathit{L}$. As shown in Figure \ref{fig:main-structure}, it combines flat and tree-like designs: each \textit{layer} maps a flat key interval and each cell may store a MAP pointer to another \textit{layer}, recursively refining the mapping.

Each \textit{layer} covers a key range $[min_{layer}, max_{layer}]$ and indexes a contiguous subset of segments $S' = \{s_j, ..., s_{j+t}\}$ with $min_j \ge min_{layer}$ and $max_{j+t} \le max_{layer}$. Its minimum resolution is:
$$
\min|layer| = \left\lceil R(layer) /\min R(s_{i'}) \right\rceil, \quad i' \in [j, j+t]
$$

and the associated SMF becomes:
$$
\text{SMF}(key) = \left\lfloor (key - min_{layer}) / R(layer) \cdot |layer| \right\rfloor
$$

\begin{algorithm}[h]
  \caption{BuildMultiLayerLM}
  \label{alg:build-mlLM}
  \small 
  \KwIn{$S$: the set of segments, $\mathit{L}$: max layer size}
  \KwOut{$lm$: Multi-layer LM}
  $k_0 \gets$ left range bound of $S$\;
  descending sort $S$ by $R(s)$, $s \in S$\;
  $s' \gets$ pop($S$)\;
  $min\_size \gets \lceil R(S) / R(s') \rceil$\;
  $lm \gets$ an array of size $\min(\mathit{L}, min\_size)$\;
  $rate \gets$ size($lm$) $/ R(S)$\;
  \While{$min\_size > \mathit{L}$ and $S$ is not empty}{
    $pos_1 \gets $ int($(min_{s'} - k_0) \times rate$);\tcp{\small Scaling mapping}
    $pos_2 \gets $ int($(max_{s'} - k_0) \times rate$)\;
    \If{$pos_1 = pos_2$ and $lm[pos_1] = NULL$}{
    \tcp{\small Build a child layer}
        $S' \gets \{s'\}$\;
        range($S'$) $\gets$ mapping\_index($lm[pos_1, pos_2]$)\;
        $lm[pos_1] \gets $BuildMultiLayerLM($S'$, $\mathit{L}$)\;
    }
    \Else {
        draw\_multi\_layer\_map($lm$, $pos_1$, $pos_2$, $s'$)\;
    }
    $s' \gets$ pop($S$)\;
    $min\_size \gets \lceil R(S) / R(s') \rceil$\;
  }
  $S \gets S \cup s'$\;
  \tcp{\small Draw remaining segments, they don't build new layers}
  \For{$s \in S$}{
    $pos_1 \gets $ int($(min_{s} - k_0) \times rate$)\;
    $pos_2 \gets $ int($(max_{s} - k_0) \times rate$)\;
    draw\_multi\_layer\_map($lm$, $pos_1$, $pos_2$, $s$)\;
  }
  \Return{$lm$;}
\end{algorithm}

Algorithm \ref{alg:build-mlLM} summarizes the recursive construction of the multi-layer LM. In lines 4 and 17, we compute the minimal resolution for the current segment. If it exceeds $\mathit{L}$ and this segment collapses into one cell, we recurse (lines 11–13) to build a child \textit{layer}. Recursion continues until the mapped range spans at least two distinct cells. The helper function \textit{draw\_multi\_layer\_map()} fills the mapped cells and ensures correct pointer assignment at segment boundaries.

Each child layer can refine current resolution by a factor of $\mathit{L}$, while occupying only one cell in its parent \textit{layer}, yielding a compact but arbitrarily fine-grained hierarchical mapping.

\textbf{Queries.} Querying follows the same SMF-based mapping strategy. Given a key, multi-layer LM computes its target cell via SMF. If the cell contains a MAP pointer, the query proceeds recursively to the next \textit{layer}. This process repeats until the result points to a MODEL or a NULL. Although hierarchical, the maximum layer depth is bounded by a small constant (see Section~\ref{sec:ca-lm}), ensuring an overall query cost of $O(1)$, making multi-layer LM suitable for high-performance lookup workloads.


\textbf{Insertions.} Insertions also follow SMF-based localization. (1) If a segment spans multiple cells ($\text{SMF}(min_i) \ne \text{SMF}(max_i)$), pointers are directly written to the mapped range. (2) If a segment maps to a single cell and the cell is not a MAP, LM checks whether the required resolution exceeds $\mathit{L}$. If so, it recursively constructs a child layer; otherwise, it scales the current layer by an integer multiple. (3) At the root level, when insertions extend the global key range and the layer size exceeds $2\mathit{L}$, LM triggers a rebuild to maintain compact layout.

\textbf{Deletion.} Since the query computation is independent of \textit{layer} size, LM does not shrink the \textit{layer} even when $\min R(s_{i'})$ increases due to deletions. A non-root \textit{layer} is only removed when its segment subset $S'$ becomes empty. At the root, if the total key range shrinks such that the ideal resolution falls below 25\% of the current layer size, LM triggers a rebuild to restore efficient memory utilization and stable query performance.

\subsection{Complexity Analysis of LM}
\label{sec:ca-lm}
In practice, LMIndex employs a multi-layer Linear Map. Unless otherwise noted, the following analysis assumes a multi-layer LM by default.

\subsubsection{Query Complexity}
\label{sec:lm-query-complexity}


Within each layer, a query requires only constant-time computation using the SMF. Thus, the overall query complexity is $O(t)$, where $t$ is the maximum depth of multi-layer LM. Let $\mathit{L} = 2^l$, then the depth $t$ is given by $t = \log_\mathit{L}(\mathtt{R}) = \log_2(\mathtt{R}) / l$, where $\mathtt{R} = R(S)/\min R(s_i)$. Since the key type determines a fixed bit-width $b$, $\mathtt{R}$ is naturally bounded by $m \le \mathtt{R} \le 2^b$, where $m$ is the number of segments. Note that $\mathtt{R}$ decreases as the distribution becomes more linear or as the segment ranges become more uniform. Consequently, $O(t) = O(b/l) = O(1)$, since both $b$ and $l$ are constants.

Following this logic, in theoretically, any $O(\log n)$ index is also $O(1)$ because $n \le 2^b$. However, in practice, significant distinctions exist between the LM and such traditional structures (e.g., B+Tree): First, thanks to the $O(1)$ intra-layer search cost enabled by the SMF, $l$ can be configured to be sufficiently large, thereby effectively reducing the theoretical upper bound of $t$. In contrast, B+Trees employ binary search within each node, which constrains the node size (fan-out) to maintain efficiency. Second, the depth of the LM is independent of the dataset size, whereas the depth of a B+Tree scales strictly with that size. Notably, when $n$ approaches $2^b$, the data becomes highly linear, causing the LM depth to converge to its minimum. Finally, the above derivation assumes each segment contains only a single key. In reality, \cite{pgm20} proves each segment contains at least $2\varepsilon$ keys, resulting in a smaller depth.

For instance, with $\mathit{L} = 65536$ (i.e., $2^{16}$) and standard 64-bit keys, the depth is bounded by $t \le 64 / 16 = 4$. Empirical results (see Table \ref{tab:lm_info}) further confirm that the actual depth $t$ remains small in practice.

While increasing $\mathit{L}$ reduces $t$, excessive values yield diminishing returns. Larger values of $\mathit{L}$ may increase space consumption, harm cache locality, and degrade update efficiency. Importantly, $\mathit{L}$ serves as an adaptive upper bound rather than a fixed capacity, allowing LM to adjust layer sizes dynamically according to segment distribution. In practice, setting $\mathit{L} = 65536$ achieves a favorable balance between query efficiency and space overhead.

\subsubsection{Update Complexity}

In the single-layer LM, the cost of updating a segment $s_i$ is determined by the number of cells it occupies:
$$
\text{SMF}(max_i) - \text{SMF}(min_i) \le R(s_i) / R(layer) \cdot |layer| + 1
$$


For a segment with the minimum key range length $\min R(s_i)$, the number of occupied cells is constant. Thus, the update cost is $O\left(\min R(s_i) / R(layer) \cdot |layer|\right) = O(1)$. Consequently, the update cost for a general segment $s_i$ is quantified as $O(R(s_i) / \min R(s_i))$.

Considering amortized complexity across $m$ segments, we obtain:
$$
O\left(\frac{1}{m} \sum_{i=1}^{m} R(s_i) / \min R(s_i) \right) = O\left(\text{avg}(R(s_i))/\min R(s_i)\right)
$$

Hence, the more the width of the segment tends to be uniform ($\text{avg}(R(s_i))\xrightarrow{}\min R(s_i)$), the closer the amortized cost is to $O(1)$.



In the multi-layer LM, updating a segment requires first locating the target \textit{layer} in $O(t)$ time. Within that layer, updates affect at most $O(\mathit{L})$ cells, following the same logic as the single-layer case. Since segment boundaries may span multiple layers, updates may involve up to $O(t\mathit{L})$ cells. Based on Section \ref{sec:lm-query-complexity}, and setting $\mathit{L}$ large enough depending on the key type, we get:
$$
O(t + \mathit{L} + t\mathit{L}) = O(t\mathit{L}) = O(1)
$$


Layer allocation or release is triggered when the current layer fails to assign at least one cell to a new segment, or when the deletion of an existing segment renders the layer empty. While this process may cascade, it affects at most $t\mathit{L}$ cells. In practice, most updates modify only a few cells. Layer allocation is infrequent, while lazy reclamation further amortizes the overhead of release.

LM's rebuilds are only triggered when the root layer becomes highly unbalanced, e.g., if the size of the root layer is exceeds $2\mathit{L}$ or the theoretical resolution is 25\% of the current size. Since each rebuild exponentially adjusts $R(layer)$, the probability of repeated rebuilds rapidly decreases.

Notably, LM does not require rebuilding to maintain correctness. It maintains robust query and update performance under distribution shifts, tolerating accumulated updates via bounded space redundancy rather than accuracy loss. Stability experiments in Section~\ref{sec:stability} confirm this resilience.


\subsubsection{Space and Construction Complexity}

In multi-layer LM, each segment occupies at most $O(t\mathit{L}) = O(1)$ space under a fixed key type. Given that there are $m$ segments in total, the overall space complexity of LM is $O(mt\mathit{L}) = O(m)$. In fact, $O(t\mathit{L})$ is quite loose. Table~\ref{tab:lm_info} demonstrates that the average number of cells per segment is far smaller than the theoretical peak value $t\mathit{L}$. Additionally, since LM requires no model retraining, its construction overhead primarily consists of pointer allocation. Consequently, the construction complexity aligns with the space complexity, both being $O(m)$.

To understand how space grows with $\mathit{L}$, we consider the worst-case bound $t\mathit{L} = \mathit{L} \cdot \log_\mathit{L}\mathtt{R}$.

Where $\mathtt{R} = R(S) / \min R(s_{i})$ is regarded as a constant related to the dataset, we compute the derivative:
$$
\frac{d}{d\mathit{L}} \left(\mathit{L} \cdot \log_\mathit{L} \mathtt{R}\right) = \ln \mathtt{R} \cdot \frac{\ln \mathit{L} - 1}{(\ln \mathit{L})^2}
$$

So when $\ln \mathtt{R} > 0$, this function achieves its minimum at $\mathit{L} = 3$ ($\mathit{L}$ is an integer), representing the theoretical space-optimal max layer size. For $\mathit{L} > 3$, the function increases monotonically. 


In practice, space usage is closely tied to the model distribution. Due to LM’s adaptive layer allocation, larger values of $\mathit{L}$ can sometimes reduce space overhead. For instance, consider $9$ adjacent segments of equal range: when $\mathit{L} = 3$, they require $12$ cells, while setting $\mathit{L} \ge 9$ reduces the requirement to $9$ cells.

\section{Indexing within a Segment}
\label{sec:single-model}
This section describes how LMIndex performs indexing within a segment, focusing on the OETA and the strategy for error correction.

\subsection{Optimal Error Threshold Algorithm (OETA)}
\label{sec:oeta}

OETA is built upon Piecewise Linear Approximation (PLA) \cite{line81}, which fits a linear model $pos = m \cdot key + b$ to a set of points such that the maximum prediction error for each point is bounded by the threshold $\varepsilon$. In practice, we typically normalize $key$ as $t$ by subtracting the first key of the segment, i.e., $t_k = key_k - key_0$, to ensure that $t_k$ is non-negative. For the $k$-th point $(t_k, pos_k)$, the set of all parameters $(m, b)$ in the dual space that satisfy the $\varepsilon$-constraint forms a strip bounded by two parallel lines:
\begin{equation}
\begin{aligned}
&\text{Lower Bound: }b \ge -t_k \cdot m + (pos_k - \varepsilon) \\
&\text{Upper Bound: }b \le -t_k \cdot m + (pos_k + \varepsilon)
\end{aligned}
\end{equation}

The intersection of all such strips forms a convex hull $P$ (specifically, a convex polygon), representing all feasible parameters. The upper boundary of $P$ consists of the tightest segments from the upper bounds of the intersecting strips, and the lower boundary is similarly derived from their lower bounds. A new segment is triggered in PLA when the strip of an incoming point no longer overlaps with $P$.

OETA calculates the minimum error threshold $\varepsilon_i$ for segment $i$ based on its convex hull $P_i$. Let $\varepsilon_i = \varepsilon - \Delta\varepsilon$. The $\varepsilon$-constraint requires that there exists $(m, b) \in P_i$ such that:
$$-t_{low} \cdot m + pos_{low} - \varepsilon + \Delta\varepsilon \le b \le -t_{up} \cdot m + pos_{up} + \varepsilon - \Delta\varepsilon$$

For a fixed $m$, $\Delta\varepsilon$ is maximized when the boundaries meet:
$$-t_{low} \cdot m + pos_{low} - \varepsilon + \Delta\varepsilon = -t_{up} \cdot m + pos_{up} + \varepsilon - \Delta\varepsilon$$

Solving for $\Delta\varepsilon$ yields:
$$
\begin{aligned}
\Delta\varepsilon &= \frac{1}{2} \left[ (-t_{up} \cdot m + pos_{up} + \varepsilon) - (-t_{low} \cdot m + pos_{low} - \varepsilon) \right] \\
&= \frac{1}{2} (f(m) - g(m))
\end{aligned}
$$

where $f(m)$ and $g(m)$ are the upper and lower boundary functions, respectively (both are linear piecewise functions). Since $f(m)$ - $g(m)$ represents the vertical chord length of the convex hull, maximizing $\Delta\varepsilon$ is equivalent to finding the \textbf{Maximum Vertical Chord Length} of $P_i$. Due to the opposing curvatures of the upper (concave) and lower (convex) boundaries, the Vertical Chord Length function $L(m) = f(m) - g(m)$ is inherently unimodal. Consequently, $\max L(m)$ can be efficiently determined by a linear time scanning algorithm that identifies the ``slope flip vertex'', which is the first vertex (from left to right) where the slope difference of the incident edges to the right satisfies $f'_+(m) \le g'_+(m)$, signifying the transition from divergence to convergence of the boundaries.

\begin{algorithm}[t]
  \caption{OETA}
  \label{alg:oeta}
  \small 
  \KwIn{(\textit{upper}, \textit{lower}): sets of convex hull vertexes (except for the upper left corner point $L$ and the lower right corner point $R$), $\varepsilon_{global}$: a global error threshold}
  \KwOut{$(m, b, \varepsilon)$: optimal segment parameters}
  $(m_p, b_p) \gets (0, 0)$\;
  $A \gets $\textit{upper} $\cup$ \textit{lower}, sorted by increasing $m$\;
  \For{$a \in A$}{
    \If{$f'_+(m_a) \le g'_+(m_a)$} {
        $(m_p, b_p) \gets a$\;
        \textbf{break}\;
    }
        
    
  }
  \tcp{\small calculate maximum $\Delta\varepsilon$ and corresponding parameters}
  $\varDelta\varepsilon \gets \frac{1}{2} \times\max(b_p - g_+(m_p), f_+(m_p) - b_p)$\; 
  $b \gets b_p = f_+(m_p) $ ? $b_p - \varDelta\varepsilon : b_p + \varDelta\varepsilon$\;
  $\varepsilon \gets \lceil \varepsilon_{global} - \varDelta\varepsilon \rceil $\;
  \Return{$(m_p, b, \varepsilon)$;}
\end{algorithm}

Algorithm \ref{alg:oeta} implements this process: Lines 3–6 scan vertexes in increasing order of $m$. Except for the upper left corner point $L$ and the lower right corner point $R$, which are the intersection points of the upper and lower boundaries. Line 4 quickly detects the slope flip vertex by slope comparison. Lines 7-9 calculate the minimum error threshold and corresponding model parameters.

\textbf{Complexity.} The dominant overhead of OETA lies in traversing the vertices of the convex hull. For a segment containing $N$ points, the resulting convex hull consists of at most $O(N)$ vertices. We demonstrate that reaching this worst-case upper bound imposes strict constraints on the data distribution. In practice, however, the empirical complexity on real-world datasets is significantly lower than this theoretical limit.

\begin{lemma}
\label{lemma:oeta}
    Under the dense key layout ($pos_{i+ 1} - pos_i = 1$), for every data point $p_i$ to contribute a unique vertex to the upper boundary or the lower boundary of the feasible convex hull $P$, the sequence of key intervals $\{\Delta t_i\}$ ($\Delta t_i = t_{i+1} - t_i$) must be strictly monotonically increasing or decreasing.
\end{lemma}

\textbf{Proof.} We focus the proof on the upper boundary of the feasible hull $P$ (the proof for the lower boundary follows a symmetric logic). In the dual space, the upper constraint for the $i$-th data point is defined by the strip upper bound line: $f_i(m) = -t_i \cdot m + (pos_i + \varepsilon)$. Since the normalized keys $t_i$ are strictly increasing, the slopes of these lines, $-t_i$, are strictly monotonically decreasing.

The upper boundary of the convex hull $P$ corresponds to the lower envelope of the set of functions $\{f_i\}$, defined as $E(m) = \min_i \{f_i(m)\}$. A specific line $f_i$ constitutes a distinct segment of this envelope (i.e., contributes a vertex) if and only if there exists a non-empty open interval $(m_{start}, m_{end})$ where $f_i(m)$ is strictly smaller than all other lines in the set.

Due to the monotonicity of the slopes, the valid interval for $f_i$ is bounded on the left by the intersection with the preceding line $f_{i-1}$ and on the right by the intersection with the succeeding line $f_{i+1}$. Let $m_{u, v}$ denote the $m$-coordinate of the intersection point between lines $f_u$ and $f_v$, $m_{u, v} = \frac{pos_v - pos_u}{t_v - t_u}$. For the interval $(m_{i-1, i}, m_{i, i+1})$ to be non-empty (i.e., $m_{start} < m_{end}$), the necessary and sufficient condition is
$m_{i-1, i} < m_{i, i+1}$. Substituting the $m_{u, v}$ formula and applying the dense layout constraint ($pos_{i+1} - pos_i = 1$), the inequality simplifies to:
$$\frac{1}{\Delta t_{i-1}} < \frac{1}{\Delta t_i} \xrightarrow[]{\Delta t_i > 0} \Delta t_i < \Delta t_{i-1}$$

Thus, the sequence of key intervals $\{\Delta t_i\}$ must be strictly monotonically decreasing. (Conversely, maintaining all vertices on the lower boundary of $P$ would require $\{\Delta t_i\}$ to be strictly monotonically increasing).

\begin{table}
  \centering
  \caption{Statistics of each segment $\max(|upper|, |lower|)$ (except for $L$, $R$) after bulk loading, $\varepsilon$=64}
  \label{tab:oeta}
  \begin{tabular}{ccccc}
    \toprule
     &\textbf{uni}&\textbf{books}&\textbf{osm}&\textbf{fb}\\
    \midrule
    \textbf{Avg} & 1.012 & 1.031&1.113 & 1.075\\
    \textbf{Max} & 3 & 7 & 11 & 9\\
  \bottomrule
\end{tabular}
\end{table}

In practice, the accumulation of a large number of vertices is statistically rare due to two dominant factors: local $\Delta t$ oscillation and curvature limits. First, real-world data rarely maintains the strict $\Delta t$ monotonicity as lemma \ref{lemma:oeta}. A single significant oscillation point (e.g., a local ``reverse bend'' in the key trend) generates a steeper strip then preceding points and typically cuts across the existing hull boundary more aggressively, thereby rendering the intermediate bounds redundant and excluding them from the active boundary. Second, segments with high curvature will exhaust the fixed error threshold $\varepsilon$ quickly. Consequently, such segments are forced to terminate while the number of data points $N$ is still small, which naturally limits the total vertex count. 

As a result, most segments fall into two categories: long and linear or short and curved, both of which result in a low vertex count. As evidenced in Table \ref{tab:oeta}, the $\max(|upper|, |lower|)$ remains approximately 1.0 across all datasets, with a maximum observed value of only 11. In summary, due to the above constraints and the inherent random entropy of real-world data, the practical overhead of OETA approaches $O(1)$, a characteristic further manifested in the experimental results in Section \ref{sec:oeta_vs_ls}.

\subsection{Choosing the Error Correction Algorithm}
\label{sec:choose-correction}
To guarantee correct query results, learned indexes typically perform error correction after model prediction. Among the commonly adopted correction algorithms, binary search and exponential search are the two dominant options. Their efficiency depends on prediction accuracy and data distribution characteristics.

Let $d$ denote the average prediction error. Binary search operates within a fixed correction range and performs independently of $d$, offering stable and predictable cost. In contrast, exponential search adaptively expands its search radius without a predefined bound, making it efficient when predictions are extremely accurate but causing its cost to grow rapidly with larger $d$.

Let $\hat{d} = 2^{\lceil \log d \rceil}$. A common implementation of exponential search first performs $O(\log \hat{d})$ exponential probes to determine the key's range, followed by a final binary search within the range of size approximately $\hat{d}/2$. Thus, the execution time for exponential search is $T_{\text{exp}} = C_{\text{exp}}(\log \hat{d} + \log (\hat{d}/2)) = C_{exp}\log (\hat{d}^2 / 2)$, and binary search with a predictable time $T_{\text{bin}} = C_{\text{bin}} \log \varepsilon$. Due to exponential search contains binary search, typically, $C_{\text{exp}} \geq C_{\text{bin}}$. Even assuming $C_\text{exp} \approx C_{\text{bin}}$ through aggressive engineering, the the runtime ratio $T_{\text{exp}}/T_{\text{bin}} = \log(\hat{d}^2/2)/\log\varepsilon$ exceeds 1 when $\hat{d}^2/2 > \varepsilon$. Figure \ref{fig:oeta_vs_ls} presents the empirical average error in complex or skewed real-world datasets, showing that $\hat{d}^2/2 \gg \varepsilon$.
\begin{table}
  \caption{Performance Comparison of Hybrid Search vs. Exponential Search in LMG across datasets from Section \ref{sec:evaluation}}
  \label{tab:hs_vs_es}
  \begin{tabular}{cccccc}
    \toprule
     \textbf{global $\varepsilon$}&32&64&128&256&512\\
    \midrule
    \textbf{Avg} & +4.71\% & +3.25\%& +0.75\%& +4.89\%&+1.29\%\\
    \textbf{Max} & +19.26\%& +7.97\%& +8.08\%& +12.71\%& +4.88\%\\
  \bottomrule
\end{tabular}
\end{table}

OETA amplifies the advantage of binary search by computing the optimal error threshold $\varepsilon_i$ for each segment. In locally linear regions with small prediction error, this tightening yields binary search performance comparable to exponential search. Table \ref{tab:hs_vs_es} reports the percentage improvement in query latency achieved by hybrid search over exponential search across different global $\varepsilon$, showing the constant factor optimization provided by OETA and hybrid search further exploits the optimization space of error correction.

Furthermore, the adaptive correction strategy ensures robustness across distributions. For near-linear regions, tuning $\varepsilon_i$ limits the overhead of binary search; for highly non-linear regions, binary search guarantees bounded worst-case cost. This distribution-aware adaptability is one of the key factors that enables LMG to maintain generality and stability in practice.

\input{09_Evaluation}

\input{10_related_work}

\section{Conclusions}
\label{sec:conclusions}


This paper introduces a lightweight decoupled model-indexing structure and the Optimal Error Threshold Algorithm (OETA) as foundational mechanisms for highly efficient learned indexing. By integrating these core components, we propose LMG, a robust learned index framework. Extensive evaluations across six practical dimensions demonstrate that LMG mitigates inherent performance tradeoffs, maintaining efficiency and multi-dimensional stability.

\bibliographystyle{spmpsci}      

\bibliography{cleaned_references}   

\end{document}

%% file: 01_intro.tex
\section{Introduction}
\label{sec:intro}
The rapid development of information technology has resulted in an enormous amount of data, posing significant challenges to data management systems \cite{bigdata16}. Indexes are vital components that enhance query performance by avoiding exhaustive searches. Traditional indexes \cite{btree06, btree11, bwtree13, dynamichash98, hash16, lsmtree96, lsmtree17, remix21, adaptive18, adaptive22}, such as B+Tree, do not take full advantage of data distributions, leading to inefficiencies caused by pointer chasing and cache misses \cite{learned23}. Learned indexes \cite{case18}, proposed in recent years, regard index as a model of data distribution, and makes proper use of this model to improve query efficiency and reduce index size.

In recent years, learned indexes have emerged as a compelling alternative to traditional structures by treating the task of locating a record as a prediction problem. However, accurate position prediction is extremely difficult and expensive, instead, most learned indexes predict within an error threshold, then perform a bounded search around predicted position to locate the true position. To keep both training and prediction costs low, they employ simple models, such as linear model or a single-layer neural network. Moreover, to improve overall accuracy and reduce correction overhead, the data is partitioned into multiple segments, each trained with its own model so as to capture local distributions with high precision. This segmented modeling both raises prediction accuracy and enhances scalability, enabling learned indexes to handle large, complex datasets. Empirical studies \cite{sosd20} have shown that even a simple piecewise linear model works effectively on the majority of real-world datasets. Accordingly, most existing learned indexes \cite{pgm20, radixspline20, fiting-tree19, alex20, madex20, coax23, swix24, disklearned24} adopt piecewise linear regression as their prediction backbone.

Despite these gains, some evaluations \cite{sosd20, micro22, gre22, learned23} have shown that existing learned indexes still have a series of flaws. For example, \cite{case18, plex21, towards21} is only suitable for read-only workloads; \cite{xindex20, lipp21, finedex21, chameleon24} deliver excellent point queries throughput but suffer on range queries; \cite{pgm20} performs well after bulk loading but lack stability after updates; \cite{lipp21} achieves efficient queries and updates, but incur high space usage; \cite{alex20, swix24} impose slow bulk‐load times. Critically, most proposals focus on optimizing only one or two metrics at a time, making it difficult to achieve a multidimensional trade‑off that outperforms the classic B+Tree. This gap leaves learned indexes largely experimental: selecting the appropriate index for an unknown or evolving workload is itself a challenge. As a result, learned indexes struggle to gain traction in practical, real‑world settings.

To address these limitations, we propose LMIndex, a foundational structure optimized for query and space efficiency, and its variant LMG, which employs localized gap allocation to absorb dynamic updates and ultimately achieve multi-dimensional performance balance. Their design stems from the optimization of two fundamental indexing phases: (1) the segment index step, which locates the appropriate data segment. To guarantee structural robustness under dynamic workloads, this top-layer structure needs to maintain a small memory footprint, execute theoretical $O(1)$ routing, and eliminate cascading maintenance overheads. (2) the localized correction step, which searches within the target segment to find the exact record position. To sustain high-speed retrievals without sacrificing overall system balance, this process demands lower prediction errors and optimized local search algorithms. Concretely, LMIndex and LMG meticulously co-design the routing logic and local correction mechanisms to navigate inherent structural trade-offs, leading to the following contributions:


\textbf{(1) Robust and Efficient Segment Index Structure.} To optimize the segment index step, we introduce Linear Map (LM). Compared to other segment index structures, LM eliminates the need for complex model training. Instead, it utilizes a deterministic mapping approach that flexibly and efficiently adapts to changes in the underlying segment distribution. Consequently, given a fixed key type, LM achieves $O(1)$ time complexity for both queries and dynamic updates. Empirically, it maintains robust performance across diverse datasets while incurring an average space overhead of only 4\%.

\textbf{(2) Optimal Error Threshold Compression Algorithm for Local Correction.} To optimize the local correction step, we introduce the Optimal Error Threshold Algorithm (OETA). Compared to the least squares method, OETA minimizes the maximum prediction error within each segment while achieving performance that is independent of distribution and nearly constant in time. Furthermore, we demonstrate that binary search (BS) outperforms exponential search (ES) on complex distributions. OETA accelerates BS, enabling it to match the performance of exponential search even on highly linear distributions.


\textbf{(3) Robust and Efficient Framework for Multi-Dimensional Balance.} By integrating the LM structure with the OETA algorithm, we first construct LMIndex, a foundational learned index framework employing a dense data layout. To unleash its potential in dynamic environments, we further develop LMG, a variant employing a gapped data layout and a novel gap allocation strategy. LMG mitigates the performance trade-offs inherent in state-of-the-art baselines, delivering a highly robust and efficient framework that maintains performance stability and competitive read/write throughput across all evaluation dimensions.

The remainder of this paper is organized as follows. Section \ref{sec:back&moti} describes the background and motivation. Section \ref{sec:overview} provides an overview of the overall design of LMIndex and LMG, including basic operations. Section \ref{sec:lm} introduces the Linear Map in detail. Section \ref{sec:single-model} introduces Optimal Error Threshold Algorithm. Section \ref{sec:evaluation} presents experimental analysis. Section \ref{sec:extend} explores the extensibility of our framework to broader database operations. Section \ref{sec:rel-work} reviews related work in the field. Finally, Section \ref{sec:conclusions} concludes the paper.

%% file: 02_background_motivation.tex
\section{Background and Motivation}
\label{sec:back&moti}
\subsection{The Idea of Learned Index}
The concept of learned indexes is inspired by machine learning: an index can be viewed as a model that maps keys to storage addresses, and building an index is analogous to training that model\cite{case18}. Concretely, the index model is the empirical cumulative distribution function (ECDF) of dataset, so the core idea of learned indexing is to construct models that fit the ECDF. In practice, however, fitting a complex real-world ECDF with a single global model is difficult and brittle under distribution shifts. Therefore, most practical learned index designs adopt multi-model hierarchies: an upper layer directs queries to leaf layer models, which are typically piecewise functions that provide more accurate local fits. In addition, indexes applies the final “last-mile” search corrects any residual prediction error when necessary.

One representative realization of this multi-model hierarchies is the Recursive Model Index (RMI) framework \cite{case18}. In RMI, each layer consists of multiple models, where a higher-level model selects the next-layer model, and the leaf-layer model predicts the position within its local range, followed by a correction search to find the exact key. Error correction in multi-model hierarchies can be performed either at internal/leaf nodes \cite{pgm20, xindex20, finedex21} or only at leaf nodes \cite{case18, alex20}; all corrections introduce runtime overhead. LMIndex achieves perfect accuracy in internal node predictions, eliminating the need for internal node correction and its associated cost.

At the leaf level, learned indexes often adopt piecewise linear models due to their structural simplicity and sufficient fitting capability \cite{sosd20}. Constructing a piecewise linear function typically involves two phases: segmentation and training. Some indexes \cite{alex20, swix24, carmi22} perform segmentation using independent heuristics and then train per-segment models via least squares regression. An alternative family of approaches \cite{fiting-tree19, pgm20, bourbon20} follows an $\varepsilon$-constrained construction: during segmentation, an error threshold $\varepsilon$ is enforced so that each segment’s model satisfies the constraint:

\begin{constraint}[$\varepsilon$-constraint]
\label{cons:epsi}
    For all keys $k$ in a segment, have $|\hat{p}(k) - p(k)| \le \varepsilon$, where $\hat{p}(k)$ is the predicted position and $p(k)$ is the true position.
\end{constraint}

Several methods, including RS \cite{radixspline20} and LIPP \cite{lipp21}, use greedy one-pass segment generation algorithms to produce $\varepsilon$-constrained partitions efficiently. The PGM-Index \cite{pgm20} adapts the streaming algorithm from \cite{line81} to construct the optimal piecewise linear approximation model (Opt. PLA-model), which generates the minimum number of $\varepsilon$-constrained segments in one pass. Opt. PLA constructs, for each prospective segment, a convex region of slope–intercept pairs that satisfy the $\varepsilon$-constrain and simply chooses the central point as the model parameter.


Compared to independent segmentation-then-train pipelines, $\varepsilon$-constrained approaches typically run in $O(n)$ time for segmentation, and the part of training cost is implicitly embedded in the segmentation process, yielding higher construction efficiency. However, Opt. PLA-model focuses solely on optimal segmentation and does not optimize the training phase. LMIndex builds upon this foundation with the Optimal Error Threshold Algorithm (OETA), which efficiently computes per-segment linear parameters that minimize the segment’s maximum prediction error. Combined with binary search for final error correction, OETA allows LMIndex to adapt each segment to its local distribution and significantly reduce last-mile correction overhead.

\begin{table*}[htbp]
\centering
\caption{Impact of Design Strategies on Key Evaluation Dimensions. Note that Negative/Positive denote relative impacts within the same main strategy. Unmarked entries depend on specific implementations (e.g., query performance of hybrid strategies depends on the additional structure).}
\label{tab:strategy_dimensions}
\renewcommand{\arraystretch}{1.3}
\begin{tabular}{ll cccccc}
\toprule
\textbf{Main Strategy} & \textbf{Sub-Strategy} & \textbf{Bulk Loading} & \textbf{Point Queries} & \textbf{Range Queries} & \textbf{Update} & \textbf{Stability} & \textbf{Space Usage} \\
\midrule

\multirow{2}{*}{Model training} 
& Top-Down & \nega & \pos & & \nega & \nega &\\
& Bottom-Up & \pos & & & \pos & \nega &\\
\midrule

\multirow{2}{*}{Segments indexing} 
& RMI-style & & \pos & & \nega & \nega & \pos \\
& Hybrid & & & & \pos & \pos & \nega\\
\midrule

\multirow{3}{*}{Error management} 
& All Node Error & & \nega & & & & \pos \\
& Leaf-only Error & \nega & \pos & & & & \\
& Zero Error & & \pos & \nega & & & \nega \\

\bottomrule
\end{tabular}
\end{table*}

\subsection{Design Strategy Analysis of Learned Index}
Although learned indexes demonstrate performance potentials exceeding traditional counterparts in specific scenarios, their robustness faces severe challenges under uncertain operating environments. Existing studies \cite{pgm20, lipp21} tend to highlight a small number of metrics (eg, lookup speeds) in specific scenarios, but their performance frequently decays when triggered by shifts in data distribution, hardware resource constraints, or workload fluctuations. The core pain point lies in the instability of performance or even unavailability (e.g. out of memory in some devices), which limits the potential for learned indexes to be widely deployed in practice.

The underlying cause of this fragility is that existing designs often pursue specialized optimization of single or few performance dimensions, neglecting the complex interference among different dimensions. Notably, the performance dimensions of an index, such as point query latency, space usage, and update latency, are not independent but tightly coupled. For instance, strategies that minimize point query latency to the extreme, such as LIPP \cite{lipp21}, incur a multiplicative expansion of the memory footprint compared to other baselines\cite{learned23}. This observation suggests that a lack of holistic consideration regarding these inter-dimensional trade-offs leads to the observed lack of robustness. Understanding the trade-offs of existing strategies is essential to enhance index robustness. Therefore, we present the following design strategies:

\subsubsection{Model Training Strategies}
In the context of model training paradigms, Top-Down and Bottom-Up strategies exhibit distinct trade-offs across multiple performance dimensions. (1) \textbf{Top-Down strategies} \cite{case18, alex20, swix24} treat the dataset as a whole and attempt to fit a global model. This method recursively partitions the dataset into subsets to improve precision of models. While this utilizes global distribution of the dataset to reduce random memory accesses and optimizes average read latency in static settings, the training process requires multiple passes over the full data, resulting in low efficiency during bulk loading. Furthermore, the advantages of this approach primarily stem from partitioning based on the perception of global distribution. Consequently, when the data distribution undergoes changes, either frequent cascading updates are required to maintain high efficiency, leading to significant update cost, or distribution drift is permitted, which results in a subsequent decline in query efficiency. (2) \textbf{Bottom-Up strategies} \cite{radixspline20, pgm20, finedex21}, represented by Piecewise Linear Approximation (PLA) \cite{pgm20}, greedily construct segments (i.e. data partitions containing the model) based on a strict maximum model prediction error threshold and build segment indexes upward from these segments. These methods are typically efficient for bulk loading, as they require only a single pass over the data. Furthermore, they remain insensitive to changes in the global distribution, given that the impact of insertions is usually confined to local regions. However, the fixed parameters of such methods (e.g. maximum error threshold) are immutable during construction, which may generate a large number of segments on skewed datasets. Furthermore, updates potentially exacerbate this increase in the segment count. Thus, segment indexing becomes critical and some indexing strategies may lead to severe degradation of query efficiency after updates \cite{pgm20, learned23}.

\subsubsection{Segments Indexing Strategies}
Consistent with the above, a segment is defined as a bottom data partition and its corresponding model (In context of B+Tree, it's leaf node). Similarly, in the design of indexing structures for managing segments, a significant conflict exists between structural flexibility and maintenance overhead. (1) \textbf{RMI-Style strategies} \cite{xindex20, chameleon24, sali23} index segments using models of varying precision. A request initially enters the lowest precision model and progressively predicts and transitions to the next layer of higher precision models until it is routed to the target segment. The advantage of this approach lies in the small memory footprint of model parameters and typically low prediction overhead. However, it faces a challenge in balancing precision and flexibility. High-precision models are tightly coupled with the data distribution, where structural modification operations (SMOs) may trigger cascading retraining and result in significant overhead. Conversely, models prioritizing flexibility permit prediction errors and rely on local correction, which compromises query efficiency. (2) \textbf{Hybrid strategies} employ additional structures or their model-integrated variants to locate segments, including B+Trees \cite{fiting-tree19, learnedkv24}, flat arrays \cite{swix24, uplif25}, hash tables \cite{lits24, lead26}, etc. The SMO of this approach generally exhibit less dependence on the data distribution, thereby partially mitigating the cascading retraining issues associated with previous methods. Furthermore, different underlying structures provide distinct advantages. For instance, B+Trees offer efficient construction alongside relatively balanced update and search performance, whereas hash tables support high-speed insertions and point queries. However, compared to RMI-Style strategies, hybrid approaches introduce additional space overhead. Furthermore, these designs inherit the specific limitations of their chosen structures, such as the search latency inherent in trees or the poor range query efficiency of hash tables.

\subsubsection{Error Management Strategies}
The error management strategy determines how the gap between the predicted position and the true position is bridged, triggering a triangular game among query latency, construction cost, and spatial continuity. Depending on the source of the error, it can be divided into three classes: (1) \textbf{All Node Error strategies} \cite{plex21, hlihp24} permits prediction inaccuracies across all nodes. This strategy benefits from simpler models, such as basic linear regression, and results in the minimal storage overhead for model parameters. However, error correction costs accumulate at every node along the query propagation path, which leads to reduced query efficiency. (2) \textbf{Leaf-only Error strategies} \cite{alex20, lits24, alt-index25} restricts prediction inaccuracies to leaf nodes (i.e., segment nodes), thereby reducing the frequency of error correction to $O(1)$ complexity and effectively decreasing query overhead. Achieving this strategy typically requires complex node splitting logic, which prolongs bulk loading time. Furthermore, certain implementations \cite{alex20} lack theoretical constraints on the maximum error of leaf nodes. (3) \textbf{Zero Error strategies} like LIPP \cite{lipp21} force physical positions to match predictions. While this eliminates correction search and optimizes point queries, it necessitates reserving massive empty slots. This results in spatial inflation and memory discontinuity, which severely degrades the performance of range queries.

In summary, most existing learned indexes represent compromises on specific performance metrics under limited assumptions. This design philosophy limits their generality in complex, dynamic environments. To this end, our goal is not to pursue single-point extremism but to design a robust and efficient framework. We aim to coordinate the aforementioned conflicts to achieve an optimal balance among critical performance dimensions, including bulk loading efficiency, point and range query performance, update throughput, stability, and space usage.

\subsection{Strategies Selection for Multi-dimensional Performance Balance and Efficiency}
As summarized in Table \ref{tab:strategy_dimensions}, existing strategies inherently incur trade-offs across different evaluation dimensions. With the objective of multi-dimensional comprehensive optimization, the key idea for selecting strategies is to exploit their inherent strengths while mitigating their respective weaknesses. Based on this systematic analysis, we identify two core philosophies: \textbf{(1) simple and universal strategies generally outperform complex designs tailored for single-dimensional gains; (2) Allowing constrained deviations or limited chaos is often more effective across multi-dimensions than strategies of absolute restriction or complete freedom.} Specifically, we select:

\textbf{Model training strategies.} We adopt Bottom-Up strategies. Compared to Top-Down strategies, which suffer from low bulk loading efficiency due to multiple data scans and high sensitivity to distribution drift, Bottom-Up strategies are more simple and universal. This approach requires only a single pass to complete construction, and its greedy algorithm offers the advantages of high efficiency and local decoupling. Furthermore, to address the issue of immutable preset error thresholds, we propose the OETA algorithm to compress the threshold of each segment to its minimum. To overcome the potential drawback of excessive segment generation, we introduce a novel segment indexing strategy that ensures the theoretical maximum routing length remains strictly bounded and small.

\textbf{Segments indexing strategies.} We adopt Hybrid strategies. Specifically, we design a novel flat/tree-like hybrid structure named Linear Map (LM) to index segments and mitigate the inherent drawbacks of Bottom-Up strategies. Compared to traditional structures, LM exploits segment distribution information. When the key type is fixed, the theoretical complexity of both queries and updates remains $O(1)$. The core model of this structure is essentially a derivation of the segments, enabling LM to achieve precise routing to target segment without the need for model training. This ensures simultaneous efficiency in both construction and query execution. Furthermore, experimental results demonstrate that its space consumption is second only to RMI-Style strategies.

\textbf{Error management strategies.} We adopt Leaf-only Error strategies. In practice, error management strategies can be viewed as a spatio-temporal trade-off. While the All Node Error strategies maximizes space savings at the expense of query efficiency, the Zero Error strategies achieves theoretically optimal query efficiency but it needs to allocate empty slots in advance. This requirement leads to both severe space inflation and degraded performance for range queries. The Leaf-only Error strategy represents a compromise between these two extremes. By incorporating OETA to ensure that leaf node errors remain controlled, it avoids space wastage while simultaneously maintaining high performance for both point and range queries. Furthermore, we design a gaps allocation strategy and hybrid search to further reduce the cost of error correction within leaf nodes.

%% file: 09_Evaluation.tex
\section{Evaluation}
\label{sec:evaluation}


\subsection{Experimental Setup}
All code is implemented in C++ (available at \url{github.com/Nuiqv/LMIndex}). Our single-threaded experiments are conducted on an Ubuntu 24.04 machine with an AMD R9 7945HX 2.5GHz CPU and 64GB RAM, compiled with gcc -O3 -march=native.

\subsubsection{Baseline} We evaluate LMIndex and LMG against five representative baselines. Shared parameters such as the error threshold are aligned for fairness and others remain default to simulate real-world scenarios that are difficult to tune. (1) Dynamic PGM Index (DPGM) \cite{pgm20} is the update-capable variant of PGM, using the same error threshold~$\varepsilon$ as LMIndex. (2) ALEX~\cite{alex20} adopts gapped arrays with model-based insertion and exponential search. We set its max node size equal to our max layer size $\mathit{L}$. (3) SWIX~\cite{swix24} is optimized for streaming data with a flat routing layer and strong random access support. We use default settings. (4) STX B+Tree~\cite{stxbtree19} serves as the traditional baseline, evaluated with its default configuration. (5) LIPP \cite{lipp21, learned23} adopts a Zero Error strategy, which eliminates the need for error correction search steps during the query process. LMIndex is configured with a max layer size $\mathit{L}=2^{16}$, initial error threshold $\varepsilon=64$. LMG inherits the parameters of LM without employing an additional buffer; instead, it utilizes a gapped array with an expansion rate of $\tau=0.5$ to accommodate updates. This configuration is a balanced trade-off, and parameter sensitivity is discussed in Section~\ref{sec:param-eval}.

\subsubsection{Datasets}
All experiments are conducted on four representative datasets: a synthetic dataset with uniform distribution and three real-world datasets from the SOSD benchmark~\cite{sosd20}. The Uniform dataset (\texttt{uni}, 200M keys) contains keys drawn from a globally and locally linear distribution. The Amazon Books dataset (\texttt{books}, 200M keys) exhibits global non-linearity with partially linear local distributions. The Open Street Map dataset (\texttt{osm}, 200M keys) is both globally and locally non-linear. The Facebook user IDs dataset (\texttt{fb}, 200M keys) is mostly linear but includes a small number of large outliers. All keys are of the $64$-bit unsigned integer type and contain no duplicate values, consistent with prior studies or the open-source codebases of learned baselines.

\subsubsection{Workloads}
We evaluate our approach primarily based on the following three categories of workloads. First, we consider static (read-only) workloads. For these scenarios, we perform read-only operations both immediately after bulk loading the entire dataset and after a specific volume of insertions. Second, we examine mixed read-write workloads, with the insertion frequencies set to 20\%, 40\%, 60\%, and 80\%, respectively. In these experiments, we initially bulk load 50\% of the data and subsequently execute 100M random read-write operations. Third, we evaluate range queries with lengths varying from $10^0$ to $10^5$. For these queries, we select 10K random starting keys and report the average latency. We evaluate LM exclusively under static scenarios. In addition, for point queries, we select 10M random keys and measure the latency.

\subsubsection{Metrics}
We measure time using average operation latency, recorded at nanosecond resolution with C++ \texttt{std::chrono}. Hardware-backed timers minimize instrumentation overhead to less than 5\%, rendering the interference with experimental results virtually negligible \cite{paoloni2010, najafi2021systems}. For memory, we track historical physical peak usage \texttt{VmHWM}. Unlike theoretical size, VmHWM represents the minimum spatial resources required for operation throughout the entire lifecycle. To ensure accuracy, all memory measurements subtract the initial usage before bulk loading.

\subsection{Evaluation of Diverse Dimensions}
We evaluated six key performance dimensions: bulk loading, point queries, range queries, stability, space usage, and updates. These dimensions characterize the fundamental lifecycle of an index, and critically impact its overall time-space efficiency.

\begin{figure*}
    \centering
    \begin{minipage}[b]{0.51\textwidth}
        \centering
        \includegraphics[width=1\textwidth]{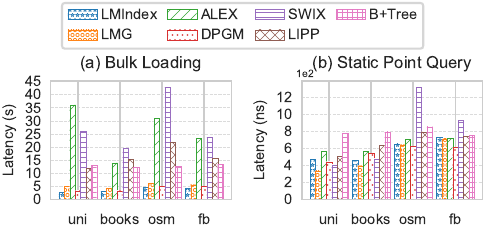}
        \caption{Bulk loading latency and static query latency.}
        \label{fig:bulk_pquery}
    \end{minipage}
    \begin{minipage}[b]{0.47\textwidth}
        \centering
        \includegraphics[width=1\textwidth]{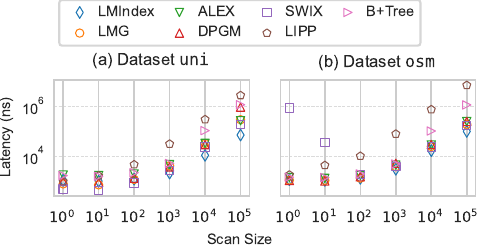}
        \caption{Static range query latency under varying scan ranges.}
        \label{fig:pure_rq}
    \end{minipage}
    \begin{minipage}[b]{1\textwidth}
        \centering
        \begin{subfigure}[b]{0.9\textwidth}
        \centering
        \includegraphics[width=1\textwidth]{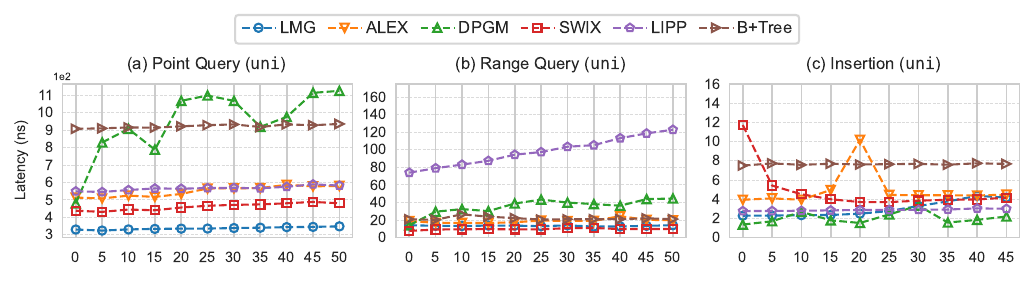}
        \end{subfigure}
        \begin{subfigure}[b]{0.9\textwidth}
            \centering
            \includegraphics[width=1\textwidth]{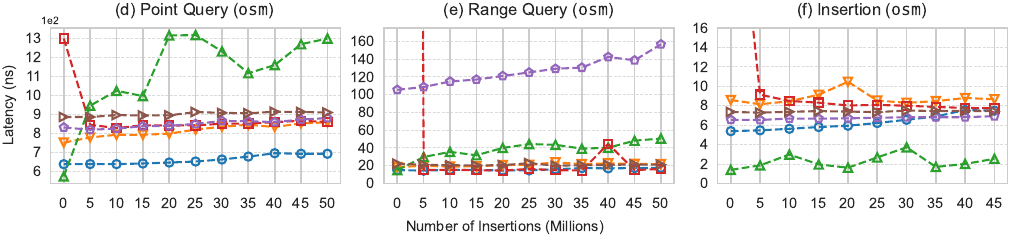}
        \end{subfigure}
        \caption{Index stability and dynamic performance during continuous insertions (0-50M).}
        \label{fig:stability}
    \end{minipage}
\end{figure*}


\subsubsection{Bulk Loading}
Bulk loading serves as a practical mechanism to initialize deployments and restore index quality after severe data distribution shifts. Figure \ref{fig:bulk_pquery}a reports the execution time across datasets, where our foundational structure, LMIndex, demonstrates superior initialization efficiency. Specifically, LMIndex achieves average speedups of 7.37$\times$ and 7.55$\times$ over top-down models like ALEX and SWIX, 4.37$\times$ over LIPP, 3.59$\times$ over B+Trees, and 1.14$\times$ over DPGM. Due to the gapped array allocation strategy, LMG slightly increases latency, being approximately 1.45$\times$ slower. Notably, the efficiency of both LMIndex and LMG does not exhibit drastic fluctuations across different data distributions.

The shared efficiency of LM and LMG inherently stems from the Bottom-Up training strategy. By greedily constructing segments through a single-pass scan over the sorted data ($O(N)$ complexity), our framework bypasses the expensive global data profiling required by other designs. Conversely, the structural bottlenecks of competing strategies become evident during initialization. Top-Down models (e.g., ALEX, SWIX) mandate recursive, multi-pass dataset scans to partition data and fit global models, resulting in significant loading delays. Similarly, Zero Error architectures like LIPP require the calculation and pre-allocation of a substantial number of contiguous empty slots, which diminishes bulk loading efficiency.

\subsubsection{Point Query}
\label{sec:point_query}
Point query execution time represents a fundamental metric for evaluating the retrieval efficiency of an index structure. Figure \ref{fig:bulk_pquery}b reports the static latency across read-only datasets. LMG demonstrates highly competitive retrieval performance, achieving average speedups of 1.30$\times$ over ALEX, 1.34$\times$ over LIPP, and 1.67$\times$ over B+Trees, while operating 1.15$\times$ faster than the base LMIndex. This baseline efficiency is primarily attributed to the Leaf-only Error strategy. By leveraging OETA to compress segment error thresholds, LMG further reduces error correction overhead, which is particularly effective on datasets with substantial latent compression potential (e.g., \texttt{uni}, \texttt{books}).

The performance differentiation becomes more pronounced under dynamic workloads with continuous insertions. As illustrated in the stability analysis (Figure \ref{fig:stability}a, \ref{fig:stability}d), competing architectures experience various degrees of performance degradation after 50M continuous insertions. For instance, DPGM, which exhibits static latency comparable to LMG, becomes 1.68$\times$ and 2.80$\times$ slower than LMG on the \texttt{osm} and \texttt{uni} datasets, respectively. In contrast, LMG sustains its point query efficiency throughout the dynamic lifecycle, extending its speedup over models like ALEX and LIPP to 1.63$\times$/1.23$\times$ and 1.68$\times$/1.28$\times$ on the \texttt{uni}/\texttt{osm}. The robustness and efficiency of point queries demonstrate that LM effectively maintains high indexing quality during updates.

\subsubsection{Range Query}
\label{sec:range_query}
In modern database systems, range queries serve as the foundational primitives for executing complex analytical operations, including relational joins and multi-dimensional aggregations (e.g., GROUP BY). Figure \ref{fig:pure_rq} reports the static range query latency across varying scan ranges (from $10^0$ to $10^5$). As anticipated, LMIndex exhibits the highest average range queries efficiency due to its strictly dense memory layout, executing slightly faster than LMG (1.13$\times$ faster). The gapped array allocation in LMG introduces a predictable, marginal overhead during sequential scans; nevertheless, LMG sustains a significant advantage over other architectures, achieving average speedups of 1.39$\times$ over ALEX, 1.30$\times$ over DPGM, and 2.54$\times$ over B+Trees. Crucially, zero error structures like LIPP incur a severe 11.41$\times$ slowdown compared to LMG. This execution penalty stems from the massive, dispersed empty slots required to enforce strict physical-to-prediction alignments, which heavily disrupt spatial locality and induce frequent cache misses during sequential traversals.

Regarding dynamic range queries, we evaluates the performance of $100$ range queries following continuous updates. As shown in Figure \ref{fig:stability}b, Figure \ref{fig:stability}e, indexes relying on fragmented experience notable performance degradation. Specifically, LIPP suffers a 7.65$\times$ slowdown on average due to progressive memory fragmentation. Similarly, DPGM (2.52$\times$ slowdown on average) experiences scan degradation because its LSM-tree style updates require traversing and merging data across multiple levels. Conversely, by leveraging efficient traversal of sequentially linked leaf nodes, B+Tree and ALEX are, on average, $1.48\times$ and $1.37\times$ slower, respectively. SWIX further strengthens its advantage through a completely flat segment indexing structure, outperforming LMG by $1.39\times$ on the \texttt{uni} dataset; however, it exhibits instability on the complex \texttt{osm} dataset, performing $1.86\times$ slower than LMG. LMG shares these structural advantages. The LM structure exhibits the characteristics of a flat array, while the segment list is ordered and linked similarly to a B+Tree. Consequently, LMG avoids severe fragmentation and the overhead of cross-level merging, ensuring both robustness and efficiency for range queries.

\begin{figure*}
    \centering
    \begin{minipage}[b]{0.32\textwidth}
        \centering
        \includegraphics[width=1\textwidth]{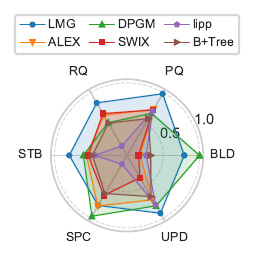}
        \caption{Comprehensive evaluation of multi-dimensional performance.}
        \label{fig:radar}
    \end{minipage}
    \begin{minipage}[b]{0.65\textwidth}
        \centering
        \includegraphics[width=1\textwidth]{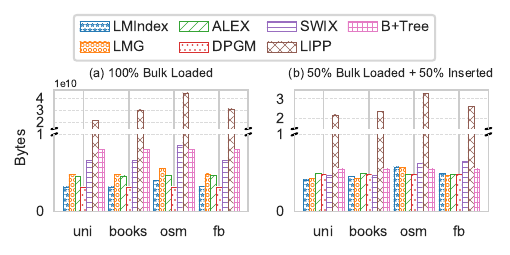}
        \caption{Space footprint comparison under static initialization versus dynamic workload conditions.}
        \label{fig:space}
    \end{minipage}
    \begin{minipage}[b]{1\textwidth}
        \centering
        \includegraphics[width=1\textwidth]{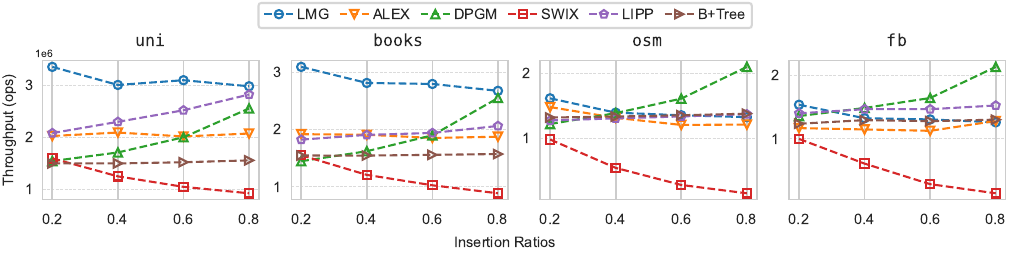}
        \caption{Overall throughput under mixed read-write workloads with varying insertion ratios (0.2 to 0.8).}
        \label{fig:update}
    \end{minipage}
\end{figure*}

\subsubsection{Stability}
\label{sec:stability}
Maintaining performance stability during continuous updates is a fundamental capability for the long-term deployment of index structures. Figure \ref{fig:stability} trace the latency evolution of point queries, range queries, and insertions across a continuous insertion workload. As analyzed in Sections \ref{sec:point_query} and \ref{sec:range_query}, LMG demonstrates sustained read efficiency throughout the operational lifecycle. From a stability perspective, this confirms the structural robustness of the LM in dynamic scenarios, which additionally benefits from the inherent robustness of the underlying hybrid search.

Focusing on the insertion operations (Figure \ref{fig:stability}c, \ref{fig:stability}f), DPGM exhibits the lowest execution latency, operating at a 1.87$\times$ faster compared to LMG. This insertion efficiency directly derives from its LSM-Tree style update strategy, which sequentially buffers incoming records without querying for positioning. However, as established in previous sections, this mechanism inherently penalizes query performance. Conversely, traditional B+Trees maintain a highly stable insertion curve due to their standardized node-splitting mechanism, yet their absolute execution time remains consistently high, averaging a 1.94$\times$ slowdown relative to LMG. Other learned baselines, including ALEX, SWIX, and LIPP, record average execution slowdowns of 1.58$\times$, 1.84$\times$, and 1.04$\times$, respectively, placing LMG's overall insertion performance in a highly competitive position.

Furthermore, it can be observed that the insertion latency of LMG gradually increases during the later stages. This trend represents an expected structural phenomenon. Initially, the gapped array allocation efficiently absorbs incoming records. As the total data volume grows and the pre-allocated gaps reach capacity, the framework increasingly triggers local segment retrains. Consequently, the frequency of these retrains naturally scales with the data volume before eventually converging to a stable rate. Nevertheless, by strictly confining these necessary retrains to localized segments, LMG successfully balances insertion costs with robust long-term read efficiency.

\subsubsection{Space Usage}
Figure \ref{fig:space} reports the historical physical peak memory usage (\texttt{VmHWM}) under static and dynamic initialization scenarios. In the 100\% bulk loaded setting (Figure \ref{fig:space}a), the spatial footprint of LMG is 1.47x that of LMIndex on average. This value is very close to the theoretical expansion of the gapped array in LMG (i.e., $(1 + \tau = 1.5)\times$). This confirms the inherent space efficiency of the base LM architecture and indicates that LMG's spatial footprint is primarily dominated by its underlying gapped array. Among other baselines, DPGM exhibits the most compact static footprint (0.63$\times$) owing to its RMI-style structure (Only the model parameters need to be stored) and dense segment layout. ALEX operates at 0.93$\times$, while SWIX and B+Trees consume 1.40$\times$ and 1.60$\times$, respectively. As established in previous sections, LIPP incurs a massive 6.26$\times$ overhead due to the physical sparsity mandated by its zero error strategy.

The spatial dynamics shift notably under continuous insertions (Figure \ref{fig:space}b). Both DPGM and LMIndex see their relative space consumption increase to 1.02$\times$ compared to LMG, highlighting their spatial instability and the fragmentation costs incurred during continuous structural modifications. Crucially, while top-down architectures like ALEX can also employ gapped arrays to absorb updates, LMG ultimately consumes less space than ALEX (1.03$\times$), SWIX (1.17$\times$), and B+Trees (1.17$\times$). This advantage stems from the inherent space efficiency of the foundational LM structure, which naturally naturally fits the underlying segment distribution and does not require a lot of routing metadata. LIPP remains highly inefficient at 5.48$\times$. Consequently, LMG's structural design effectively prevents severe memory inflation throughout its lifecycle.

\subsubsection{Update Performance}
Distinct from the stability evaluated in Section \ref{sec:stability}, this section assesses index throughput under mixed workloads, with write ratios varying from 0.2 to 0.8. Averaged across all datasets and workload configurations, LMG demonstrates highly competitive overall throughput, operating 1.31$\times$ faster than ALEX, 1.49$\times$ over B+Trees, 1.19$\times$ over LIPP, 1.26$\times$ over PGM, and 3.50$\times$ over SWIX. Specifically, LMG consistently achieves the highest throughput on the \texttt{uni} and \texttt{books} datasets, validating its efficiency under these distributions.

On the \texttt{osm} and \texttt{fb} datasets, LMG slightly trails DPGM and LIPP under specific mixed workloads. This performance variance is primarily driven by the distinct structural characteristics of the baselines. For example, the LSM-tree-style buffer in DPGM enables direct insertions without requiring query routing; however, distributing data across multiple levels degrades the efficiency of subsequent reads. Furthermore, LMG exhibits a slight throughput reduction under specific heavy-read workloads. This behavior stems from our conservative segment retraining strategy. To strictly guarantee read efficiency and preserve local error bounds, LMG proactively triggers segment retrains before the pre-allocated gaps are fully exhausted. While this deliberate architectural trade-off introduces a predictable structural overhead as write operations occur, it successfully ensures bounded, long-term operational stability without compromising read performance.

\subsubsection{Holistic Performance and Multi-dimensional Balance}
Figure \ref{fig:radar} synthesizes the holistic performance of the evaluated indexes through a normalized radar chart (score from 0.0 to 1.0). For most dimensions (e.g., Bulk Loading and Space Usage), normalized scores are derived by averaging relative performance metrics across all datasets. To accurately capture long-term index stability, the Stability (STB) dimension is quantified using the Area Under the Curve (AUC), defined as $AUC = \int_{0}^{50M} Latency(x) \,dx$. This formulation effectively penalizes unpredictable latency spikes and sustained execution delays over the continuous insertion lifecycle. Furthermore, the Point Query (PQ) and Range Query (RQ) dimensions aggregate both static and dynamic workloads from STB to reflect a comprehensive operational lifecycle.

The resulting visualization exposes the severe multi-dimensional trade-offs inherent in existing baselines. For instance, while DPGM achieves near-optimal normalized scores in Bulk Loading (1.000) and Space Consumption (0.969), its structural rigidity severely penalizes its Range Query (0.520) and Stability (0.602) performance. Similarly, zero error architectures like LIPP catastrophically compromise Space (0.141) and Range Queries (0.148) to maintain read limits. In contrast, LMG yields the most expansive and balanced performance polygon. By consistently sustaining high normalized scores across all six dimensions (ranging from 0.789 to 0.981), LMG empirically validates its core design philosophy: avoiding catastrophic degradation in any single metric to deliver robust, general-purpose efficiency across diverse and dynamic database workloads.

\begin{figure*}
    \centering
    \begin{minipage}[b]{0.5\textwidth}
        \centering
        \includegraphics[width=\linewidth]{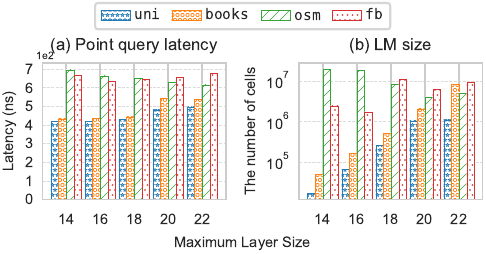}
        \caption{Evaluation of point query performance and LM size under different maximum layer sizes ($2^{14}$ to $2^{22}$) for LM.}
        \label{fig:max_layer_size}
    \end{minipage}
    \begin{minipage}[b]{0.48\textwidth}
        \centering
        \includegraphics[width=1\textwidth]{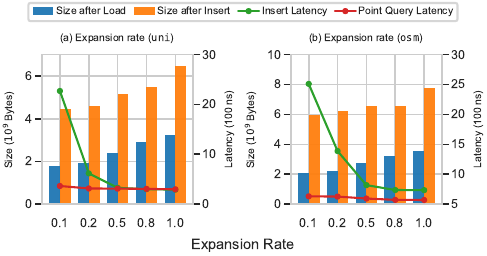}
        \caption{Impact of expansion rate on LMG in terms of point query, insert and space.}
        \label{fig:expansion_rate}
    \end{minipage}
    \begin{minipage}[b]{0.27\textwidth}
        \centering
        \includegraphics[width=\linewidth]{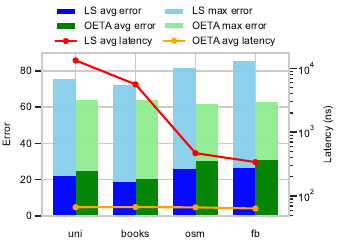}
        \caption{Analysis of OETA and Least Squares.}
        \label{fig:oeta_vs_ls}
    \end{minipage}
    \begin{minipage}[b]{0.27\textwidth}
        \centering
        \includegraphics[width=\linewidth]{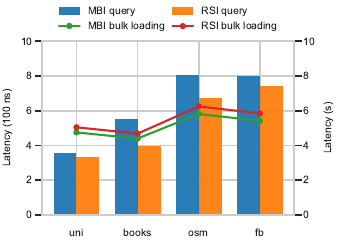}
        \caption{Comparison of gap allocation strategies.}
        \label{fig:lmg_insert_comparison}
    \end{minipage}
    \begin{minipage}[b]{0.45\textwidth}
        \centering
        \includegraphics[width=0.86\linewidth]{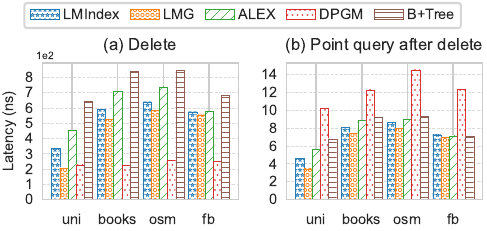}
        \caption{Deletion efficiency and its impact on query performance.}
        \label{fig:delete_test}
    \end{minipage}
\end{figure*}

\subsection{Detailed Evaluation}
\subsubsection{Parameter Evaluation}
\label{sec:param-eval}
We evaluate the impact of the maximum layer size $\mathit{L}$ on LMIndex, as well as the expansion rate $\tau$ on LMG.

\textbf{Max Layer Size.} Figure \ref{fig:max_layer_size} evaluates the sensitivity of the LM structure to its maximum layer size $\mathit{L}$ (from $2^{14}$ to $2^{22}$). Crucially, point query latency exhibits relative consistency across all $\mathit{L}$ values, demonstrating that retrieval efficiency is structurally robust and insensitive to parameter tuning. Regarding space usage, the size of LM remains relatively stable for complex data distributions (e.g., \texttt{books}, \texttt{osm}, and \texttt{fb}). Conversely, on the relatively linear and uniform dataset \texttt{uni}, expanding $\mathit{L}$ increases the number of cells per segment, leading to a significant increase in index size. However, despite this trend, the absolute LM size on the \texttt{uni} dataset remains the lowest among all evaluated distributions, and it stabilizes within the $2^{20}$ to $2^{22}$ parameter range. Overall, adjusting $\mathit{L}$ across a vast parameter space achieves relatively stable performance; a smaller $\mathit{L}$ can yield a slight advantage on more linear distributions, though the absolute magnitude of this improvement is marginal (compare to complex distributions). This further verifies the structural robustness of LM.

\textbf{Expansion Rate.} The expansion rate $\tau$ determines the proportion of reserved gaps in the data array, influencing both memory footprint and update/query performance. As shown in Figure \ref{fig:expansion_rate}, increasing $\tau$ initially reduces insertion latency, but the benefits saturates once most updates are absorbed by existing gaps. Query latency also improves slightly due to fewer correction steps, though the gain quickly saturates, especially on datasets with linear distributions like \texttt{uni}. Meanwhile, memory usage rises steadily with larger $\tau$, reflecting the overhead of excess reserved gaps. Therefore, selecting $\tau$ requires balancing update performance against memory overhead. Overall, the magnitude of $\tau$ is linearly correlated with space consumption. Regarding queries, the impact of $\tau$ remains relatively stable; for insertions, a smaller $\tau$ is preferable. However, we identify $0.5$ as a ``sweet spot'': across two datasets with vastly different distributions, it serves as the inflection point where the slope of insertion latency begins to flatten, while still maintaining a moderate spatial footprint.

\subsubsection{Gap Allocation Strategy}
We compare two gap allocation strategies within LMG: model-based insertion \cite{alex20} and reserved-space insertion. As shown in Figure \ref{fig:lmg_insert_comparison}, the reserved-space insertion incurs slightly higher bulk load time due to more complex gap placement. However, it consistently delivers better query performance by pre-allocating gaps more evenly across array. This reduces average correction length and improves search performance. Overall, given the efficient bulk loading process of LMG, the small increase in load time is an acceptable trade-off for stronger runtime performance.

\subsubsection{OETA vs. Least Squares} 
\label{sec:oeta_vs_ls}

We evaluate the training efficacy of OETA against the standard least squares (LS) baseline by measuring three key metrics: average maximum error per segment, average mean error per segment, and algorithm runtime. As shown in Figure \ref{fig:oeta_vs_ls}, OETA consistently yields tighter error bounds. Compared to the LS baseline, OETA improves the average mean error by 12.5\% on average and reduces the average maximum error by 15.4\% to 26.6\%. This mean/maximum error trade-off is particularly effective when we employ maximum error sensitive binary search. In terms of runtime, OETA demonstrates fundamental advantages in computational efficiency and stability. Experimental results indicate that OETA consumes only 0.5\% to 18\% of the runtime required by LS. Crucially, while the performance of the Least Squares (LS) method exhibits high variance, OETA maintains a nearly consistent low overhead across diverse data distributions. The experimental results manifest the characteristics of an $O(1)$ algorithm, aligning with our theoretical derivations. Furthermore, this efficiency renders OETA more suitable for time-sensitive scenarios, providing an alternative approach for computing model parameters in other learned indexes.

\begin{table}
  \caption{Statistics of Linear Map}
  \small
  \begin{tabular}{lcccc}
    \toprule
     &\textbf{uni}&\textbf{books}&\textbf{osm}&\textbf{fb}\\
    \midrule
    \textbf{Avg update latency (ns)}	& 156.86 & 159.99 & 280.13 & 111.05\\
    \textbf{Avg layer size}			& 61504	&671.6	&3333	&20.4\\
    \textbf{Avg search depth}			& 0	&0.004	&0.989	&2.184\\
    \textbf{Avg segment's cells}			& 4.7	&5.6	&135.4	&3.6\\
    \textbf{Max layer depth}			& 0	&2	&2	&3\\
    \textbf{Num of layer} 			& 1	&249	&15704	&93025\\
    \textbf{LM building time / total}	    & 0.02\% & 0.16\% & 6.16\% & 3.08\%\\
    \textbf{LM space usage / total}	    & 0.066\% & 0.17\% & 15.7\% & 1.73\%\\
  \bottomrule
  \label{tab:lm_info}
\end{tabular}
\end{table}
\subsubsection{Linear Map Evaluation} 
\label{sec:lm_eval}
Table \ref{tab:lm_info} presents key statistics of LM, except update latency measured under a 50\% random insertion workload and all other metrics collected after fully loading each dataset. The root layer is defined as depth 0. Across datasets, LM exhibits consistently low and stable update latency. Most keys are located within 1 layer, with the \texttt{fb} dataset being an exception due to the segments containing a small number of large outliers placed at the root. Owing to the flat layer structure and the training-free model design, LM achieves almost negligible build time, accounting for just 0.02\%–6.16\% of total index construction time.

\subsubsection{Delete Evaluation}
Figure \ref{fig:delete_test} evaluates the deletion efficiency by removing 50\% of the dataset and measuring the subsequent point query latency. In terms of sheer deletion speed, DPGM exhibits the lowest latency. This efficiency derives directly from its LSM-tree-style mechanism, which rapidly appends tombstone markers without verifying the actual existence of the target keys. However, this blind insertion strategy inevitably penalizes subsequent read operations, causing DPGM to perform 2.05$\times$ slower than LMG in post-deletion queries on average. Conversely, LMG employs a more balanced approach that combines in-place tombstone marking with localized segment merging and retraining. While this structural maintenance introduces a slight latency overhead during the deletion phase (though LMG still operates 1.45$\times$ and 1.85$\times$ faster than ALEX and B+Trees, respectively), it effectively bounds the deterioration of prediction accuracy. Ultimately, LMG strikes a robust architectural trade-off, efficiently accommodating massive deletions while preserving long-term query stability.

\section{Discussion and Extensibility}
\label{sec:extend}
While the current LMG architecture is fundamentally designed as a foundational framework to achieve multi-dimensional performance balance for complex workloads, its highly decoupled hybrid structure remains inherently extensible. By isolating the routing logic from the underlying physical data layout, LMG provides a versatile blueprint capable of supporting a broader spectrum of complex database operations and query paradigms.

\textbf{Duplicate Keys Support.} LMG could naturally accommodate duplicate keys by pushing the handling logic down to the localized segments. In this configuration, the upper-level LM structure would solely manage segment distribution information, remaining entirely oblivious to key duplication. During the localized model training phase, the segment would train its model exclusively on unique keys and their physical positions, thereby ensuring prediction accuracy and bounding inference errors. Consequently, point queries could efficiently locate the target primary key within the local segment, and subsequent range scans could seamlessly traverse the associated value lists without disrupting the contiguous segment alignment.

\textbf{Concurrent Operations.} To support multi-threaded database workloads, LMG could integrate standard concurrency control protocols. Because structural modifications (SMOs), such as gap allocation and segment retraining, are strictly confined to local boundaries within our hybrid architecture, LMG would be well-suited to employ fine-grained, segment-level latching (e.g., read-write locks) for the bottom data layer. This localized absorption would drastically minimize lock contention during highly concurrent insertions. Simultaneously, the upper routing layer could be protected using Optimistic Concurrency Control (OCC) or versioned locks, ensuring that read transactions traverse the hierarchy without heavy synchronization overheads.

\textbf{Vector and High-Dimensional Queries.} Although originally optimized for 1D scalar keys, LMG could be adapted to support high-dimensional vector searches. This extension could be achieved by employing order-preserving mappings, such as Space-Filling Curves (e.g., Z-order or Hilbert curves) or Locality-Sensitive Hashing (LSH), to project multi-dimensional vectors into single-dimensional scalar keys. LMG could subsequently be constructed over these mapped keys. Under this paradigm, complex high-dimensional proximity evaluations or $k$-nearest neighbor ($k$-NN) candidate retrievals would effectively be transformed into localized 1D range scans, allowing the system to leverage LMG's robust sequential scanning efficiency.

%% file: 10_related_work.tex
\section{Related Work}
\label{sec:rel-work}

\textbf{Evolution of Learned Indexes.}
The paradigm of learned indexing was pioneered by RMI \cite{case18}, which leverages the cumulative distribution function (CDF) of underlying data to map search keys to approximate physical locations. While early model-based indexes \cite{radixspline20, plex21, shift21} primarily targeted static workloads, the inherent limitations of read-only structures in dynamic environments spurred the development of updatable architectures. FITing-Tree \cite{fiting-tree19} and PGM \cite{pgm20} introduced bottom-up construction strategies governed by strict error bounds. Conversely, ALEX \cite{alex20} expanded upon the RMI concept by employing a top-down, data-driven construction that deliberately sacrifices structural balance to achieve superior model fitting across varying data granularities. These foundational works triggered a rapid proliferation of learned index designs, establishing a vast design space characterized by distinct structural layouts and optimization objectives.

\textbf{Structural Paradigms.}
The underlying routing architecture fundamentally dictates the search efficiency and update capability of a learned index. Hierarchical models typically employ either recursive, multi-layer error bounds \cite{case18, pgm20} or tree-based layouts \cite{alex20, lipp21} to iteratively refine position estimates. While tree-based designs facilitate localized updates without requiring holistic model retraining, the cascading traversal overhead remains non-trivial. Recent advancements \cite{uplif25, swix24, cabin24} advocate for flattened structures, deliberately reducing routing depth by expanding the capacity of individual layers. However, navigating these expansive layers still demands computational overhead for model inference. In contrast, our proposed LM structure addresses this bottleneck by computing deterministic mapping parameters. This enables $O(1)$ intra-layer routing without continuous model retraining, inherently supporting efficient local modifications while maintaining high accuracy.

\textbf{Model Construction and Optimization.}
Another critical dimension in learned index design is the methodology used for data segmentation and linear approximation. Segmentation strategies broadly bifurcate into top-down \cite{case18, alex20, swix24, carmi22} and bottom-up \cite{fiting-tree19, pgm20, bourbon20} frameworks. Top-down methodologies optimize segments holistically; for instance, SWIX \cite{swix24} utilizes second-order derivatives to enforce tighter linearity within segments. Conversely, bottom-up algorithms generate data-driven partitions governed by strict error thresholds. Within these segments, the regression techniques vary significantly: several frameworks optimize for average error using least squares \cite{alex20, sindex20, finedex21, swix24}, while others compute bounding parameters via convex hull algorithms to guarantee maximum error bounds \cite{pgm20, madex20, towards21}. Model retraining also incurs significant tail latency, a challenge that several studies \cite{chameleon24, vega25, hire25} have attempted to mitigate. Building upon these foundations, LMIndex leverages a bottom-up segmentation strategy coupled with the OETA algorithm to efficiently derive parameters that strictly minimize the maximum error per segment. Furthermore, this can be further refined through advanced parameter tuning \cite{tuning22, airindex23, newtuning25} and dynamic statistics maintenance \cite{sali23}.

\textbf{Expanding Modalities and Hardware Utilization.}
Beyond one-dimensional numeric keys, the learned index design space has diversified to accommodate complex data types. Researchers have tailored models for multi-dimensional vectors \cite{flood20, Tsunami20, coax23, learnedaknn22}, string data \cite{sindex20, rss21, lits24}, spatial datasets \cite{ifx20, lisa20, wisk23}, and streaming windows \cite{flirt23}. Concurrently, orthogonal optimizations have emerged to exploit modern hardware topologies, including CPU cache optimization \cite{carmi22}, GPU acceleration \cite{hyper24}, concurrent execution \cite{xindex20, finedex21}, and persistent memory efficiency \cite{apex21, simple23, learnedpmem23, disklearned24, wipe24, making24}. Despite these specialized advancements, balancing high-dimensional performance with robust, dynamic updatability remains an ongoing challenge in the field.